**Concurrent observations at the magnetic equator of small-scale irregularities and large-scale depletions associated with equatorial spread F**

Dustin A. Hickey, Center for Space Physics, Boston University, Boston, MA

Carlos R. Martinis, Center for Space Physics, Boston University, Boston, MA

Fabiano S. Rodrigues, Center For Space Sciences, The University of Texas at Dallas, Richardson, TX

Roger H. Varney, SRI International, Menlo Park, California, USA

Marco A. Milla, Jicamarca Radio Observatory, Lima, Peru

Michael J. Nicolls, SRI International, Menlo Park, California, USA

Anja Strømme, SRI International, Menlo Park, California, USA

Juan F Arratia, Ana G. Mendez University System, Puerto Rico, USA

Corresponding author: D. A. Hickey, Center for Space Physics, Astronomy Department, Boston University, 725 Commonwealth Ave, Boston, MA, 02215, USA (dahickey@bu.edu)

**Abstract**

In 2014 an all-sky imager (ASI) and an Advanced Modular Incoherent Scatter Radar consisting of 14 panels (AMISR-14) system were installed at the Jicamarca Radio Observatory. The ASI measures airglow depletions associated with large-scale equatorial spread F irregularities (10's-100's km), while AMISR-14 detects small-scale irregularities (0.34 m). This study presents simultaneous observations of equatorial spread F (ESF) irregularities at 10-100 km scales using the all sky-imager, at 3 m scales using the JULIA (Jicamarca Unattended Long-term Investigations of the Ionosphere and Atmosphere) radar, and at 0.34 m scales using the AMISR-14 radar. We compare data from the three instruments on the night of 20-21 August, 2014 by locating the radar scattering volume in the optical images. During this night no topside plumes were observed, and we only compare with bottomside ESF. AMISR-14 had five beams perpendicular to the magnetic field covering ~200 km in the east-west direction at 250 km altitude. Comparing the radar data with zenith ASI measurements, we found that most of the echoes occur on the western wall of the depletions with fewer echoes observed the eastern wall and center, contrary to previous comparisons of topside plumes that showed most of the echoes in the center of depleted regions. We attribute these differences to the occurrence of irregularities produced at sub-meter scales by the lower-hybrid-drift instability. Comparisons of the ASI observations with JULIA images show similar results to those found in the AMISR-14 and ASI comparison.

**Index Terms:** 2439, 2411

**Keywords:** ionosphere, irregularities, ESF, radar, imaging, depletion

## 1. Introduction

Equatorial spread F (ESF) is the name commonly given to plasma irregularities that typically occur after sunset in the equatorial and low latitude F region. They are attributed to plasma bubbles, depletions in the background plasma that begin to form due to the generalized Rayleigh-Taylor instability (*Hysell*, 2000). These plasma bubbles contain structures that cover many different scale sizes. Radar systems pointed perpendicular to the magnetic field are able to look at coherent backscatter from field aligned irregularities with scale sizes that are equal to half the wavelength of the radar. Observations of ESF with radar are generally categorized into three distinct types of features: bottom-type, bottomside, and topside [*Woodman and La Hoz*, 1976]. Their names refer to where they appear relative to the F-layer. Bottom-type layers are narrow layers of irregularities that do not show much vertical development. Bottomside irregularities are also limited to a small layer but are more structured and developed. Topside irregularities are plumes that reach up to the topside. At Jicamarca, Peru, an incoherent scatter radar (ISR) operating at 50 MHz has been widely used to observe ESF irregularities at 3 m scale size [*Woodman and Hoz*, 1976; *Fejer et al.*, 1999;]. A handful of radars at different longitudes are used to study ESF: the ALTAIR radar (9°N, 167° E, 155.5 MHz and 415 MHz) [*Tsunoda*, 1980b] and the Christmas Island radar (2° N, 202.6° E, 50 MHz) in the Pacific Ocean [*Miller et al.,* 2010]; the São Luís radar (2.59° S, 44.21° W, 30 MHz) in Brazil [*Rodrigues et al.,* 2008] and the Equatorial Atmosphere Radar (EAR) (0.2° S, 100.32° E, 47.0 MHz) in Indonesia [*Otsuka et al.,* 2004]. These radars detect irregularities with scale sizes from 0.36 m to 5 m.

Studies of ESF irregularities at sub-meter scale size are very scarce. Tsunoda [1980b] used the ALTAIR radar operating in the 415 MHz mode, and TRADEX (1320 MHz), concurrently on 17 July 1979. The ALTAIR radar operated in continuous E-W scans and when irregularities were detected, the TRADEX radar was turned on. They found 11 cm irregularities at the same time as the 36 cm irregularities. The 11 cm irregularities are approaching the electron gyroradius and both irregularity scale sizes are below the oxygen gyroradius. Tsunoda [1980b] claimed that these irregularities would most likely only exist in steep gradients in low-density plasma as discussed by Huba and Ossakow [1981]. At scales sizes less than 1 m, the lower-hybrid-drift instability may be responsible for the irregularities which is different from the process that creates irregularities at the meter scale [*Huba and Ossakow*, 1981]. The gradient drift instability, where the plasma drift and plasma density gradients are antiparallel, has been proposed as the source of 3 m scale size irregularities on the western wall of large-scale depletions and in the center of the depletion [*Miller et al.*, 2010].

Conventional radar observations used to observe ESF have limitations since they only provide spatial information in range (or altitude). As the irregularities pass over the radar line-of-sight they are evolving, so it limits what can be determined and one must be careful of the slit-camera effect [*Hysell*, 1996]. JULIA is a particular radar configuration used to detect coherent backscatter from plasma irregularities. One of its modes, the imaging mode, uses radar interferometry to resolve structures within the radar beam [*Hysell*, 1996; *Hysell and Chau*, 2006]. This mode can help resolve some of the spatial and temporal ambiguity but it is limited to a small area on the sky of about 6°. The imaging mode data has been compared with plasma density data from the Communications Navigation Outage Forecast System (C/NOFS) and the echoes seem to be correlated with depletions on the order of about 1 km [*Hysell et al.*, 2009].

In 2014 an Advanced Modular Incoherent Scatter Radar (AMISR) was installed at the Jicamarca Radio Observatory. This system has only 14 panels (a full AMISR system has 128 panels). The smaller number of panels means that the system has less power and a smaller antenna aperture and cannot be used for incoherent scatter radar observations. The half power beam width of the AMISR-14 system is 8° in the E-W direction and 2° in the N-S direction. This system can be used for coherent backscatter from field-aligned plasma irregularities associated with ESF. This system has an operating frequency of 445 MHz so the backscatter that is detected is from irregularities with scale sizes of 0.34 m, about an order of magnitude smaller than those detected with the Jicamarca radar. One of the major advantages of an AMISR system is that it has an electronically steerable beam. This allows measurements at multiple positions on the sky almost simultaneously. This feature is useful for observing larger portions of the sky at once.

Rodrigues et al. [2015] presented the first results of ESF detection with AMISR-14 at Jicamarca. They were able to use the multiple beams of AMISR-14 to observe bottomside irregularities moving in the eastward direction with an average velocity of about 113 m/s, and covering over 200 km in zonal distance at times. The variation and decay of the echoes can be observed as they pass through the beams. During the same nights that the campaign was being run, the Jicamarca radar was also running in the JULIA mode. They concluded that both radars observed irregularities near the same time and altitudes, although some differences were detected and attributed to differences in the observational modes and hardware (e.g. antenna beam width, power, etc.).

These two radar systems are able to observe small-scale irregularities present in ESF. One way to probe much larger scale sizes, 10's -1000's km, is to use an all-sky imager (ASI). This has previously been done away from the magnetic equator, where the features map to higher altitudes at the magnetic equator [*Martinis and Mendillo*, 2007]. The ASI installed at the Jicamarca Radio Observatory in March 2014 is used to measure large-scale irregularities simultaneously with the Jicamarca ISR and AMISR-14.

The comparison between observations of ESF made, simultaneously, by radars and ASIs had yet to be done at Jicamarca. The Equatorial Atmosphere Radar (EAR) was used with an ASI to study from one night in April, 2003 when there were field aligned irregularities occurring within the airglow depletions. The area with the most intense backscatter was found to be in the center of the depletion [*Otsuka et al.,* 2004]. Mendillo et al. [2005] showed the process of mapping airglow depletions along magnetic field lines and showed a case study of simultaneous radar backscatter and optical data. Miller et al. [2010] conducted a similar study using an ASI in Hawaii looking south and the Christmas Island radar, which also observes 3 m irregularities. They also found that the backscatter at high altitudes comes from the center of the depletions. Since none of these studies were done at the magnetic equator, in order to compare the data from both instruments, they had to project the radar data onto the ASI images. This projection is done assuming that the irregularities will map perfectly along the field lines. The detected echoes were from high altitude plumes (h> 400 km).

In this paper we present observations of ESF over Jicamarca on the night of 20-21 August 2014 with the two radars and the ASI. We are able to observe ESF with irregularities at 0.3 m and 3 m scale sizes, using the radars, and at larger scale sizes by measuring airglow depletions.

## 2. Data and Methods

### 2.1 Radar and optical diagnostics

The ASI at the Jicamarca Radio Observatory can be used to observe the mesosphere and ionosphere through narrow band interference filters designed to detect airglow emissions in these regions. A background filter is also used for calibration purposes. We are interested in emissions at 630.0 nm and discuss only this filter in this study. The filter has a full width at half maximum of about 1.0 nm. The optical assembly limits the angle of incidence on the filter to less than about 4° to the normal. The filter transparency varies by no more than 10% within this angle and the greatest deviation is only from the edge of the images at 90° viewing angles (Baumgardner et al. 1993). This emission is limited to an altitude range of about 50 km and typically occurs at 250 km altitude. Light at 630.0 nm is produced by dissociative recombination of molecular oxygen ion, followed by the de-excitation of atomic Oxygen and is dependent on the neutral and plasma densities as well as on the height of the ionospheric layer. In an all-sky image of the sky, a weak emission means that there is a decrease in plasma density or that the ionospheric layer moved to heights of lower recombination. We are able to use this emission to observe ESF which, at the magnetic equator, appears as large-scale plasma depletions that are aligned with the magnetic field and move across the field of view from west to east. It is possible to resolve depletions as small as about 10 km but most are around 100 km in width. Small-scale structures have been observed to occur at the same time as the large-scale structures [*Miller et al.,* 2010, *Otsuka et al.,* 2004] but the connection is not fully understood.

Data from the JULIA imaging mode has been used to study ESF for many years now [*Hysell*, 2000] while the AMISR-14 system is relatively new and ESF has been reported only once [*Rodrigues et al.,* 2015]. We start by looking at the data from the JULIA imaging mode on the night of 20-21 August, 2014. The right panel in Figure 1 shows the range-time-Doppler-intensity (RTDI) map of the JULIA observations made during pre-midnight hours of 20 August. The brightness of the backscatter shows the relative intensity of the observed echoes and the color represents the Doppler velocity of the irregularities producing the echoes. Red represents

Doppler velocities away from the radar, blue represents velocities towards the radar, and green represents near zero velocities. Other colors represent a mixture of Doppler velocities. The left panel shows an example of the distribution of the irregularities causing echoes within the radar beam obtained using interferometric imaging. This particular image corresponds to measurements made around 21:44 LT (local time is UT-5). While the JULIA imaging mode provides high angular resolution images of the irregularities, it can only provide information for a field of view of a few degrees off zenith. AMISR-14 complements and expands the interferometric imaging measurements by looking at different directions farther from the zenith. We determine where in the ASI images the AMISR-14 radar beams are located and then determine which part of the depletion the radar echoes are located. We discuss this technique further in the following paragraphs in the context of the AMISR-14 comparison.

The advantage of the AMISR-14 system is that it is electronically steerable so that it can look at multiple locations on the sky with the capability of switching pointing directions from pulse to pulse. With this technique we can compare the radar data with multiple places in the ASI images and observe how the small-scale irregularities change as they move from beam to beam. This experiment used seven different look directions, five of which were perpendicular to the magnetic field in order to observe the field-aligned irregularities. The beam configuration is shown in the top panel in Figure 2. Beams 2 through 6 are the beams that are perpendicular to the magnetic field and are the only ones used in this study. This experiment used 28-bit coded pulses with a baud length of 10 µs and an inter-pulse period of 4 ms. 128 pulses were transmitted in each pointing direction before switching and 1280 pulses, in each direction, were integrated to obtain an AMISR-14 image. This means that we have an integration time of 36 seconds for the seven beams.

We use the AMISR-14 and ASI to compare backscatter from small-scale structures to the large-scale depletions. We determine where in the large-scale structures the small-scale irregularities are appearing. The 630.0 nm images are taken about every eight minutes with an exposure time of 120 seconds and since the AMISR-14 integration time is about 36 seconds there are about four radar measurements during one image. To compare between the two datasets, we choose the AMISR-14 data that is closest to the center of the exposure time from the imager. The backscatter from irregularities in the AMISR-14 data barely changes in the exposure time of the image. Three consecutive time steps are shown in at the bottom of Figure 2. There are some cases where irregularities are only present in one beam. As the irregularities pass over the radar and through the beams, they tend to occupy one beam for about five minutes. Figure 3 shows the entire night of data from the zenith beam in the same format as Figure 1.

Two types of ESF structures are observed in the measurements presented in Figure 1 and Figure 3. The first type, observed before 21:00 LT, is bottom-type ESF and is seen very often in radar preceding the detection of large-scale ESF events; it is not associated with Rayleigh-Taylor density perturbations. It is, instead, the result of interchange of plasma instabilities (wind-driven gradient drift) occurring on horizontal electron density gradients occurring around sunset. The echoes are caused by weak plasma turbulence associated with small background density perturbations. GHz scintillation, for instance, would be negligible during that type of ESF.

After 21:00 LT, we observe the passing of a bottomside ESF structure, which are irregularities associated with an under-developed plume/depletion, which was presumably created by the Rayleigh-Taylor instability process. The irregularities are in a more turbulent state and cause stronger radar echoes.

The detection of bottom-type, bottomside and topside ESF radar structures with AMISR-14 are discussed in Rodrigues et al. (2015). A description, in more depth, of these structures is given by Woodman and LaHoz (1976) and Hysell and Burcham (2002). We still seek a better understanding of the conditions and plasma instabilities causing these ESF structures, but recent improvements have been made (e.g. Aveiro and Hysell, 2010).

Overall, similar behavior of the irregularities in the two radars is observed although there are a few differences. At some altitudes, features persist for longer in the AMISR-14 data than they persist in the JULIA data. For example, between 21:00 and 21:30 LT, we see echoes for a longer time in the AMISR-14 data compared with the JULIA data. We attribute this to differences in the sensitivity of the two systems. While AMISR-14 can still observe 0.3 m irregularities, the power-aperture of the JULIA mode was not enough to detect, clearly, echoes from the 3 m irregularities. At other times the JULIA mode detects features that are not detected by AMISR-14. This can be seen when looking at altitudes higher than 400 km. The JULIA data shows echoes above 400 km but none are seen in the AMISR-14.

We next describe how the raw all-sky images are processed to be compared with the radar data. A raw image at 23:00 LT on this night is shown in Figure 4a. ESF depletions are the large dark bands that are N-S aligned. Due to distortion effects as one moves away from zenith we must first unwarp the image, i.e., map it to a certain height and assign a latitude and longitude position to each pixel. We assume an emission height of 250 km and the elevation and azimuth angles to determine the longitude and latitude of each pixel. We then overlay a map with grid lines. The unwarped image at 23:00 LT using a field of view covering a zenith angle of 80° is shown in Figure 4b. The features in the center of the image shown in Figure 4a become much smaller in the unwarped image. Stars and the Milky Way emit blackbody radiation that contains 630.0 nm so they are visible in our images. The stars have been removed from the images using an algorithm that replaces brighter pixels with the median of the surrounding pixels. The Milky Way is not removed by this technique since it is so large that removing it would affect the visibility of the depletions. The ESF is now visible as mostly straight depletions that are aligned in the N-S direction.

The contrast in the images can make it difficult to see the depletions so we take a zonal cut through the center of the image and plot pixel value as a function of distance from the center of the image. Seven pixels in the meridional direction are averaged to create the cut. There are 136 pixels for each degree of latitude. Figure 5 shows ten zonal cuts from 21:49 LT to 23:00 LT, which are the times that we are focused on in this study. We start at 21:49 because the airglow is too dim early in the night due to the F peak being too high in altitude. At 23:00 LT, six depletions are clearly visible, compared to earlier in the night when only one or two are visible. A depletion has been marked as depletion "A" and can be tracked all the way back to 21:57 LT. The depletions are moving at approximately 100 m/s, consistent with the movement of the irregularities from Rodrigues et al. (2015). The Milky Way can be seen in Figure 5 as an increase in brightness that is much smaller in zonal width than the depletions. At 21:49 it is at a zonal distance of about -70 km and moves west as the night goes on ending up at about -250 km.

**2.2 Comparing the Radar with the ASI**

Once unwarped images are produced the radar data can be compared with the ASI data. We present data from 22:13 LT on 20 August 2014 in Figure 6 to demonstrate how we perform the comparison. Figure 6a shows an unwarped all-sky image with a box in the middle representing the area of the zoomed-in region that is visible in Figure 6b. The western coast of South America is seen as a white line. The white cross in the center shows the position of the ASI and radar, which is located at -0.1° geomagnetic latitude. The two N-S white lines indicate the coverage of the 5 perpendicular to B AMISR-14 beams. They extend North and South to show how they map along magnetic field lines. Airglow structures occurring away from zenith, to the north or south, can be mapped to apex heights (i.e., height above the magnetic equator) higher than 250 km. The dashed lines indicate 275, 325, and 375 km apex heights for field lines reaching 250 km off the equator. An apex altitude of 275 km is at a magnetic latitude of ±3.5° and an apex height of 375 km is at a magnetic latitude of ±7.8°. The relatively bright airglow regions represent background plasma density and the darker regions are depleted regions or airglow depletions where there is less plasma. The transition between dark and bright regions, where the plasma density changes, is what we call the walls of the depletions. The western wall of the depletion is the transition from high plasma density to low plasma density and the eastern wall is the transition from low to high density. Different regions of the image are placed into one of four categories: the western wall, the center of the depletion, the eastern wall, and the bright background. If the beam includes parts of both walls it is categorized as either the center of the depletion or the bright background, even if one wall is more represented with the beam width. The depletions are N-S aligned and in general cover the entire field of view.

In Figure 6b we show the zoomed-in box from Figure 6a that covers a zenith angle of 50°. The locations of the AMISR-14 beams are shown in the image. This can be compared directly with the AMISR-14 data shown in Figure 6d. In the N-S direction the beam only covers 2° (8.75 km, 7 pixels). The cross in the center shows the location of the Jicamarca radar. Geodetic latitude is marked as is the W-E zonal distance. The image size was chosen to match the radar data. In this image we can see a bright area near the center and the darker depletions are on either side. The Milky Way can also be seen in the image and is distinguished from the depletions by its orientation, size, and the direction of movement.

In order to make the comparison between the imager and the radar we have to make sure that our image is unwarped at an accurate altitude. The emission mostly comes from a region that is about 50 km in altitude extent and is typically centered around 250 km. The height of emission does vary through the night. We have looked at how different altitudes could change the results and varying them by ~50 km (a typical height variation in nighttime 630.0 nm airglow) does not significantly affect the result. The dark and bright zones are still in the same beams.

As was done with Figure 5, we take a zonal cut at zenith, averaging over 7 pixels north and south. This is plotted as pixel value versus zonal distance in Figure 6c. We also do a running average in the zonal direction of 15 pixels to help see the larger features better. The running average is the smoother curve plotted over the noisier curve.

## 3. Results

We compare data from AMISR-14 with images from the ASI to determine where in the large-scale depletions the small-scale echoes are occurring. We refer to Figure 6 to demonstrate what the comparison looks like. In this example, at 22:13 LT, we see that beam 2 shows no echoes and the region corresponds to the eastern wall of the airglow depletion marked as "A" in Figure 5. In beam 3 AMISR-14 detects irregularities, still on the eastern wall. Beam 4 is on the western wall of a different depletion and echoes from AMISR-14 are also present. Beam 5 encompasses a small depletion that includes both western and eastern wall gradients and we see echoes in AMISR-14. Beam 6 is on the western wall of yet another depletion and again strong echoes are seen with AMISR-14.

Figure 7 shows the comparison for a later time, now at 22:29 LT. Here we can see that depletion "A," that was to west of beam 2, has now moved into beam 3. This depletion is approximately 200 km in width, showing the size of these large-scale structures. Beams 2 and 6 are the only ones that detect significant echoes. Beam 2 is on the western wall of depletion "A" and beam 6 encompasses a part of the bright background and a western wall and is categorized as the bright background. The center of the depleted region and the eastern wall does not show any significant echoes.

We counted all the times where we do and do not see echoes at zenith from 21:49 UT to 22:52 LT that corresponds to the times when AMISR-14 detected echoes in the 225-275 km range. We choose a 50 km range because that is approximately the altitude extent of the airglow layer. If there are echoes higher up, they are not considered and are analyzed later. Table 1 shows these results. If there are echoes that have an SNR greater than -5 dB, this situation is marked as **yes**. If there are no echoes in the beam this is marked as a **no**. For each image we look at whether the beam encompasses an eastern wall, a western wall, the background plasma or the depletion center.

We see scatter in 13 beams that correspond to the western wall, such that about 48% of cases occur here. There are no instances when we observe the western wall and do not see scatter. We see scatter in 5 beams that corresponds to the eastern wall, which is about 19%. We found that 11 beams show no scatter on an eastern wall. We also see scatter in 4 cases where there is a bright peak, which represents the background plasma. As can be seen in the zonal cuts, the background plasma peak is smaller in width than the AMISR-14 beam so these areas contain gradients as well. There are no instances when we observe a bright peak and no scatter. We see scatter in 5 cases where the beam is in the depletion center, which also often contains gradients, and 2 cases where there is no scatter and the beam is in the depletion center.

Another way of analyzing this data is to determine how frequently each section of a depletion has radar scatter relative to the amount of times that part of the depletion is in a beam. This is found by taking the number of times radar scatter is observed for a certain part of the depletion and dividing it by the number of times that section is in a radar beam. Scatter is detected in the western wall 100% of the time and in the eastern wall it is about 31%. Scatter is detected on the bright background 100% of the time and in the depletion center it is about 71%. The bright background and depletion center always contain parts of both walls which must be taken into account when interpreting the results. From this information we can conclude that scatter comes from all regions of the depletion but it is more likely to occur in the western wall. The scatter from the western wall is most likely due to the gradient drift instability [*Sekar et al.,* 2007; *Miller et al., 2010*] because the density gradient and drift direction are antiparallel on this wall. In contrast, the scatter that is found on the eastern wall seems to be due to a different process since the gradient and drift direction are not antiparallel which is also supported by the difference in the occurrence rates.

Large-scale depletions observed by the ASI allow us to compare them with echoes that are observed from altitudes > 250 km. Structures north and south of zenith tell us information about radar echoes at the apex heights marked in Figures 5 and 6.

We did a similar analysis for cases when echoes were observed between 275 and 325 km (corresponding to magnetic latitudes of ~3.5-6° in the ASI FOV) and between 325 and 375 km (corresponding to magnetic latitudes of ~6-8°). 50 km ranges were chosen because the airglow layer is approximately 50 km in altitude extent. The airglow structures seem to show a slight N-S asymmetry with respect to the magnetic longitudes so we separate these two results. As mentioned before, in the analysis between 225 and 275 km, if echoes were present above 275 km, they were not included. Now these results, as well as those with echoes from 325-375 km, are included in this analysis.

In Table 2 we show the combined results from 275-325 km and 325-375 km. Results from the north and south show the same general trend that more echoes come from the western wall than from anywhere else. If we combine the north and south results we find that about 46% of the scatter is on the western wall and ~ 32% of the scatter is on the eastern wall. The background and depletion centers do not represent 'flat' structures but still include gradients from both the western and eastern wall and ~ 23% of the scatter is seen in those as well.

We also compute the percentage of how frequently each region of the depletion has radar scatter for this two altitude ranges, as we did for 225-275 km. We find that radar scatter occurs on the western wall about 85% of the time and occurs on the eastern wall about 68% of the time. The bright background has scatter about 61% of the time and the depletion center has radar scatter about 71% of the time

The percentages for total number of occurrences are very similar to the results from 225-275 km. The percentages for chance of occurrence vary a bit when compared with the lower altitudes. The western wall still shows a high frequency of occurrence even though it is no longer 100%. The eastern wall now shows a much higher frequency. The depletion center shows about the same frequency as the lower altitudes and the bright background is still high even though it is no longer 100%. The differences between the higher altitudes and lower altitude could be attributed to the fact that we are dealing with small numbers at 225-275 km so it does not take much to vary the percentage.

We have to assume that the features map perfectly along field lines to do this analysis and this assumption also introduces some error as well. If the depletions mapped perfectly along the magnetic field lines then the magnetic field line with a foot point in the center of a depletion in the southern half of the image would have another foot point in the center of the depletion in the northern half. When we compare the location of these foot points to the north and to the south we find that the depletions to the south are often located to the west of the foot point that maps to the center of the depletion to the north. An inverted "C" structure is expected of these depletions due to the wind shear in the plasma flow that peaks near the F peak [*Kelley et al.,* 2003] but the "C" should be symmetric about the magnetic equator. The asymmetry could be due altitude and latitude gradients in the background winds. Even though the features north of zenith do not map to the exact same location at the magnetic equator as the ones south of zenith we still find the same result that echoes come from everywhere in the depletion but the western wall is favored. Overall, the results from Table 1 and 2 show that the most echo detections occur on the western wall.

The same comparison is done with the JULIA imaging mode but this radar is only able to point in one direction so we are only able to look at one part of the ASI image instead of five. This decreases the amount of comparisons that can be done. Early in the night the ESF echoes occur between 300 and 400 km, which is higher than the 630.0 nm emission at 250 km, and the echoes do not last very long after they appear at this altitude. From the images during this time we find three cases when echoes appear on the western wall of the depletion. We do not find any cases where the echoes occur in the center of the depletion, the eastern wall, or the background.

We do not compare the higher altitude JULIA imaging data with the parts of the ASI images that map to higher altitudes because the beam width is much smaller with this mode than with AMISR-14 so any errors in the mapping along field lines would be amplified. Although results are not conclusive due to the limited number of observations we see the same preference for echoes to occur at the western wall, like the AMISR-14 comparison.

In addition to the comparisons shown above we can also track irregularities across the beams very accurately to see how their evolution relates to the motion of the airglow depletions. We show in Figure 8 an example of this from 22:29 LT to 22:44 LT. Arrows indicate the echoes and the parts of the image they correspond to. We can clearly see that in all three images the echoes are on the western wall of depletion "A" as it moves through the field of view. The earliest time in Figure 8 also shows that the western wall of another depletion to the east also contains irregularities. The motion of the western wall is correlated with the motion of the irregularities. The most intense echoes on the western wall, weak echoes are sometimes appearing in the center of the depletion and no echoes are on the eastern wall. We are not able to track echoes on the eastern wall in the same way at different times because when echoes appear on the eastern wall they tend to also be occurring in all of the beams.

Our results show more echoes are found on the western wall of the large-scale airglow depletions. On the western wall of the depletion the plasma density gradient is antiparallel to the nighttime neutral wind which is the condition for the gradient drift instability. This instability has been proposed as the source of irregularities on the western wall [*Miller et al.,* 2010]. Otsuka et al. [2004] proposed that the low-frequency drift instability is responsible for the small-scale echoes they observed since the most intense echoes observed in their study occurred in the center of the depletions. Miller et al., [2010] discussed that the echoes in the center may be driven by or propagating from the echoes that occur on the western wall due to the gradient drift instability. They also observed that only the high altitude echoes (h > ~400 km) were mainly coming from the central part of the depletions.

The AMISR-14 irregularities are sub-meter in size, which means that it is likely that the lower-hybrid-drift instability is involved as well and is contributing to the echoes. There are certain conditions for this instability to exist as outlined in Huba and Ossakow [1981]. Specifically, this can occur at 0.34 m with gradient scale lengths smaller than ~50 m, for a neutral density of $10^{15}$ m$^{-3}$, and electron density of $10^{10}$ m$^{-3}$. Results from the ionosonde at Jicamarca show that the background electron density is about $10^{10}$ m$^{-3}$. Using NRLMSISE-00 [*Picone et al.,* 2002] we find that the neutral density should be around $10^{14}$-$10^{15}$ m$^{-3}$. This means that with relatively sharp density gradients, which we expect to exist in these depletions, the lower-hybrid-drift instability is likely to be occurring.

The results presented here do not seem to agree with results from Otsuka et al. [2004] and partially agree with those from Miller et al. [2010]. In these previous studies, ASIs and radars were not located at the magnetic equator. Thus they were sampling depletions at higher altitudes, most likely associated with topside plumes, whereas we are observing only bottomside ESF. The different conditions present for topside plumes compared with bottomside ESF could explain the difference in the location of the echoes. AMISR-14 is probing sub-meter scale sizes and the previous studies were looking at ~3 m scale sizes. The lower-hybrid-drift instability cannot produce 3 m scale irregularities [*Huba and Ossakow,* 1979], which can explain why our results are different.

We plan to look at more simultaneous observations of the AMISR-14 and ASI since we only present one night of data here. Other nights when AMISR-14 was running and detected ESF echoes did not have clear sky conditions to observe depletions with the ASI Additional JULIA imaging data can also been incorporated to better constrain where the echoes are coming from.

## 4. Summary

We have compared ASI observations with AMISR-14 measurements in an attempt to better understand the relative localization of large-scale (10-100s km) equatorial spread F depletions and small-scale irregularities. Overall we find that echoes occur preferentially on the western wall of the depletion over all other parts. This is supported by the comparisons between the 250 km AMISR-14 data and zenith of the ASI, between higher altitude AMISR-14 data and off zenith ASI depletions, and between JULIA imaging data and zenith of the ASI.

We find that when comparing the zenith measurements of the ASI with 250 km echoes from AMISR-14, approximately 48% of the echoes occur in the western wall of depletions, 19% on the eastern wall, and the other 33% occurs in the background and depletion center. Looking at how often echoes were detected in each part of the depletion we find that 100% of the time echoes are detected on the western wall and 31% of the time they are detected on the eastern wall. We extend this comparison to look at the airglow north and south of zenith that correspond to apex heights at the magnetic equator of 300 km and 350 km. We combined the data from the north with zenith and south with zenith and found very results. 46% of scatter is on the western wall, 32% on the eastern wall, and 23% in the bright and dark regions. We find that 85% of the time echoes are detected on the western wall and 68% of the time on the eastern wall.

We have found that there is a higher occurrence of echoes coming from regions identified as the western wall of large-scale depletions in comparison with echoes coming from other regions of the depletion. Our results suggest that the gradient drift instability is responsible for the echoes on the western wall at 0.34 m and 3 m and that the lower-hybrid drift instability is responsible for the echoes on the eastern wall at 0.34 m and may also contribute to the echoes on the western wall. The lower-hybrid-drift instability is not capable of producing instabilities at 3 m. Our results only include observations of bottomside ESF and do not show the same trends observed in previous studies of topside ESF using ASI and radar data. The previous studies analyzed 3 m scale size irregularities. We only find a difference between our results from AMISR-14 and previous studies. Our JULIA results are consistent with previous results. We attribute the differences to the fact that we are observing sub-meter scale irregularities produced by the lower-hybrid-drift instability.

We have also shown that during one part of the night we are able to track the echoes occurring on a western wall as they pass through multiple beams and during this time it is only the western wall that is producing echoes.

**Acknowledgements**

This work was supported by grants from the Office of Naval Research for DURIP instrumentation at Jicamarca Observatory and data analysis funds from the NSF's programs for Aeronomy #1123222 and # 0925893. DH acknowledges NSF grant IIA-1139862, which allowed student travel to Jicamarca. Work at UT Dallas was supported by AFOSR (FA9550-13-1-0095) and NSF (AGS-1303658). Work at SRI International was supported by NSF Co- operative Agreement AGS-1133009. Work at Anna G. Mendez University System was supported by NSF (AGS-1039593). JULIA measurements used in this study were made available by the Jicamarca Radio Observatory is a facility of the Instituto Geofisico del Peru operated with support from the NSF AGS-1433968 through Cornell University. JULIA measurements and contact information can be found at http://jro.igp.gob.pe. Information about the AMISR-14 measurements presented in this study can be obtained by contacting Fabiano Rodrigues (fabiano@utdallas.edu).

**Table 1 Cases at 250 km where radar echoes were found in each part of the depletion**

| Echoes 250 km | Western Wall | Eastern Wall | Background | Depletion Center |
|---|---|---|---|---|
| Yes | 13 | 5 | 4 | 5 |
| No | 0 | 11 | 0 | 2 |

**Table 2 Radar echoes detected in each part of the depletion for altitudes above 250 km**

|  | Echoes | Western Wall | Eastern Wall | Background | Depletion Center |
|---|---|---|---|---|---|
| South | Yes | 37 | 24 | 6 | 8 |
|  | No | 6 | 10 | 7 | 3 |
| North | Yes | 31 | 23 | 11 | 9 |
|  | No | 6 | 12 | 4 | 4 |

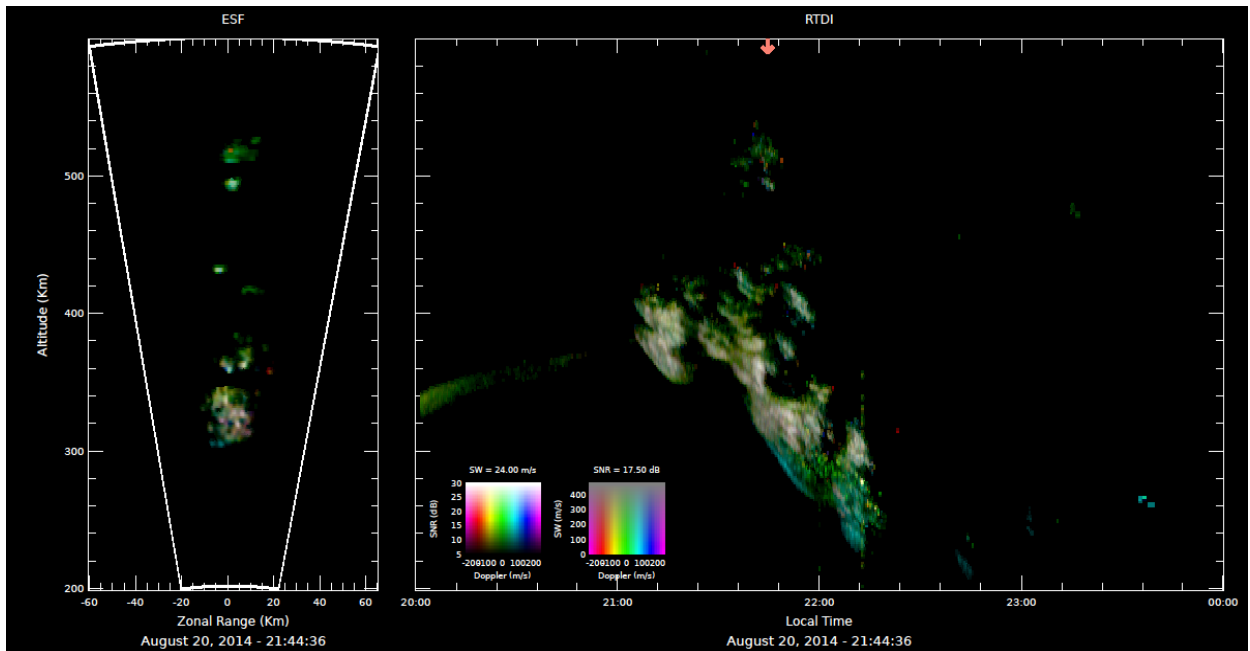

Figure 1

A range-time-Doppler-intensity plot (right) from the Jicamarca radar in the JULIA imaging mode from the night of 20-21 August 2014. The brighter regions are higher signal to noise of the echoes. The brightness of the backscatter shows the relative intensity of the observed echoes and the color represents the Doppler velocity of the irregularities producing the echoes. Red represents Doppler velocities away from the radar, blue represents velocities towards the radar, and green represents near zero velocities. Other colors represent a mixture of Doppler velocities. The plot is a combination of the individual times, an example of which is shown on the left. The time on the left is 21:44 LT

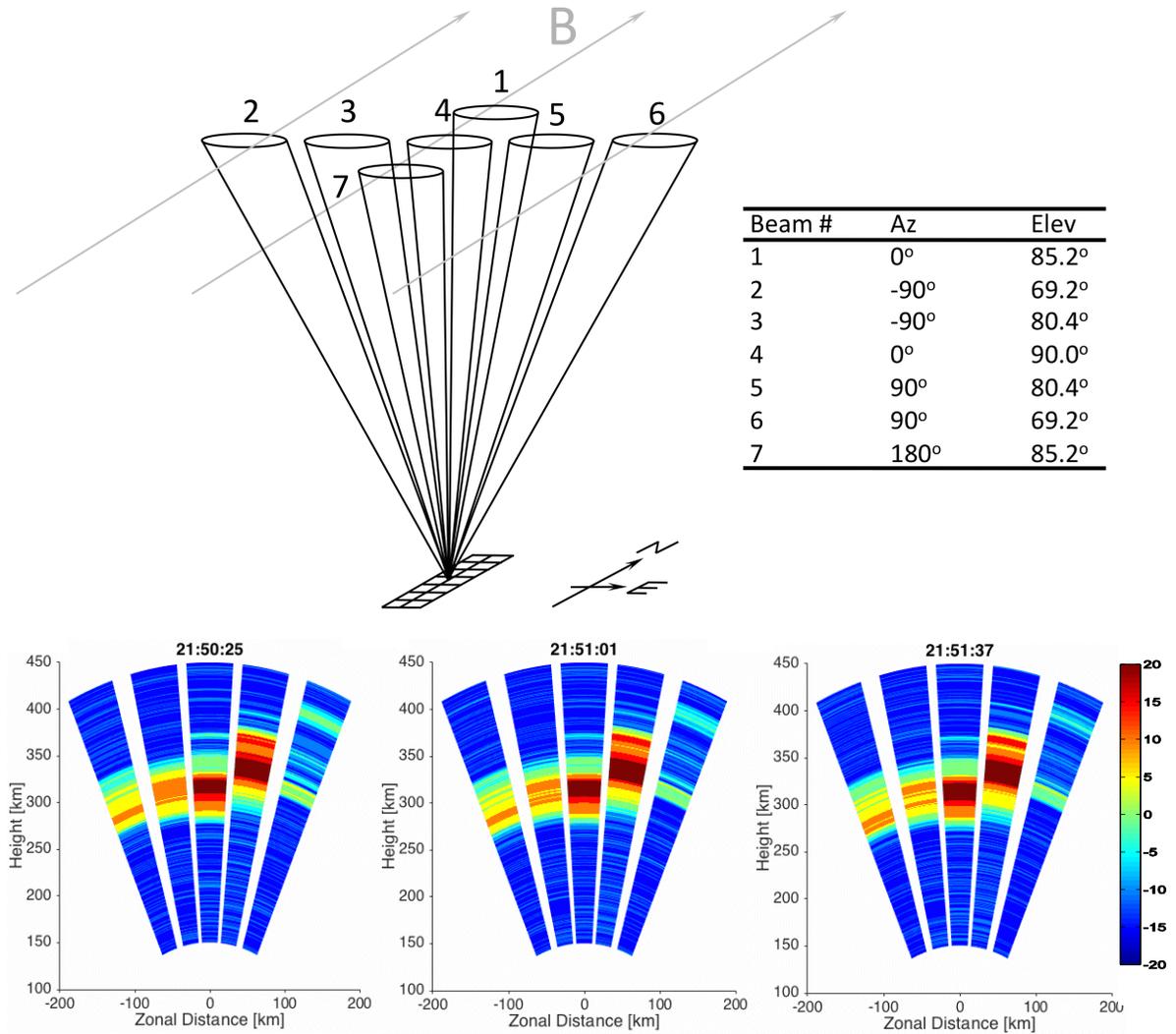

Figure 2

The top diagram illustrates the beam directions used in the AMISR-14 experiment used in this study (taken from Rodrigues et al., 2015). The bottom panels show three AMISR-14 scans obtained with the multi-beam experiment. The colors represent the intensity of the echoes in dB. The color bar to the right indicates the signal to noise ratio of the echoes in dB. This is approximately the amount of data taken during one ASI image. There is not much variation between the three times.

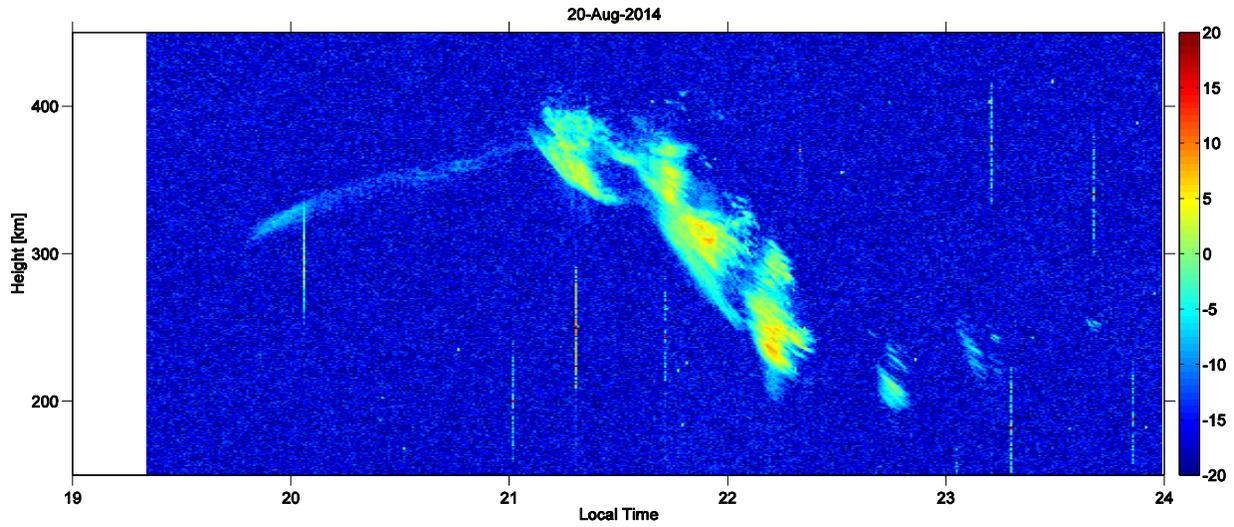

Figure 3

A range-time-intensity (RTI) map for the beam 4 (zenith) observations made by AMISR-14 on the night of 20-21 August 2014. The color bar to the right indicates the signal to noise ratio (SNR) of the echoes in dB. The vertical streaks seen throughout the night are from satellite echoes and are not related to the properties of the atmosphere.

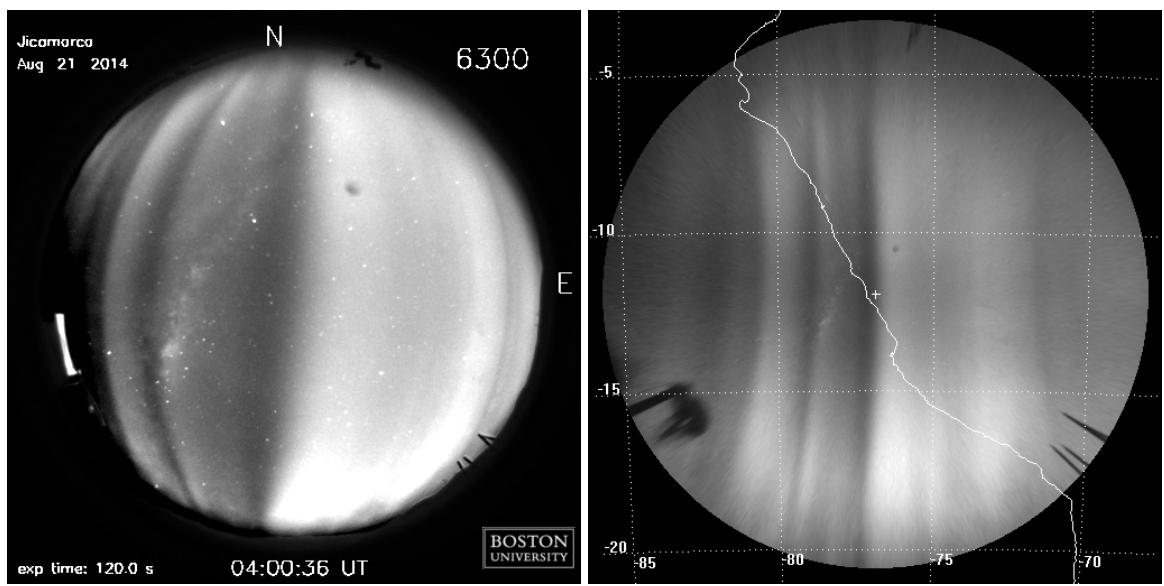

Figure 4

A raw image at 23:00 LT (4:00 UT) from the Jicamarca ASI on the left. The exposure time is 120 seconds and the image was obtained using a 630.0 nm filter. On the right is the result of unwarping the image. The western coast of South America is seen as a white line and the location of the ASI is marked with a small white cross. The ASI is located a magnetic latitude of -0.1°. The dotted lines are geographic latitudes and longitudes. Latitude and longitude is determined for each pixel and the image is transferred to a map projection.

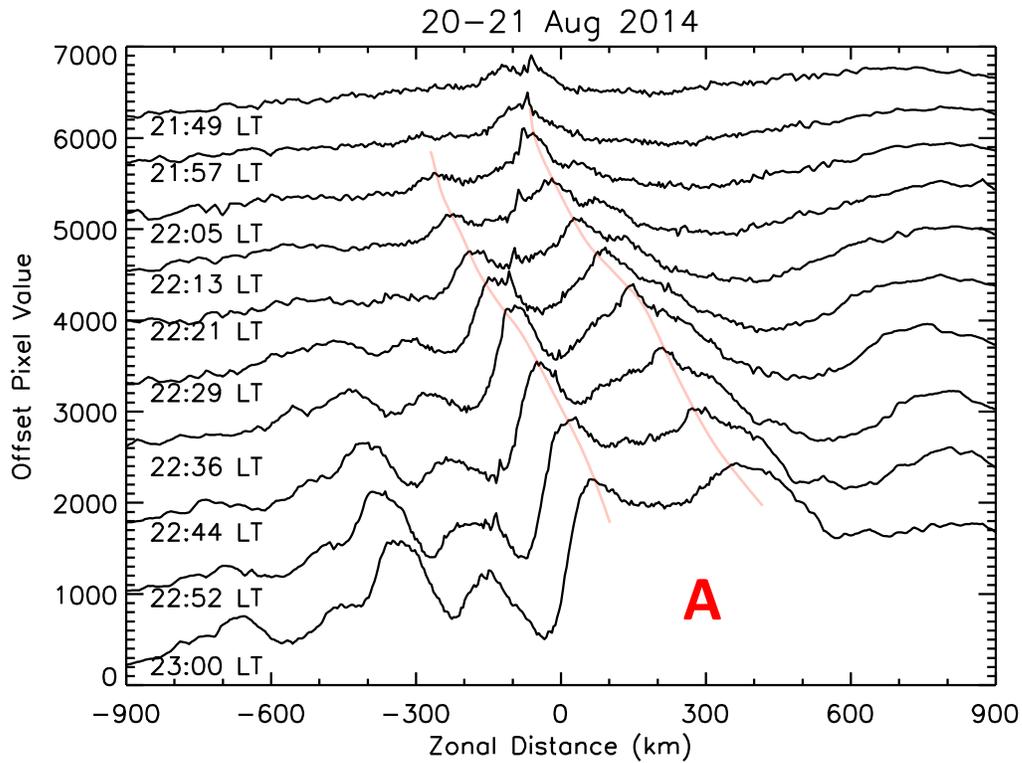

Figure 5

This plot shows the pixel value for zonal cuts for all the images from 21:49 LT to 23:00 LT. The pixel value is arbitrary as each time is offset from the others to make it easier to see. The cuts are 7 pixel meridional averages. 7 pixels correspond to the meridional beam width. The movement of the depletions across the field can easily be seen. The depletion marked as A, between ~100-300 km east at 23:00 LT, is the one we mostly focus on. The red lines follow the west and east edges of the depletion.

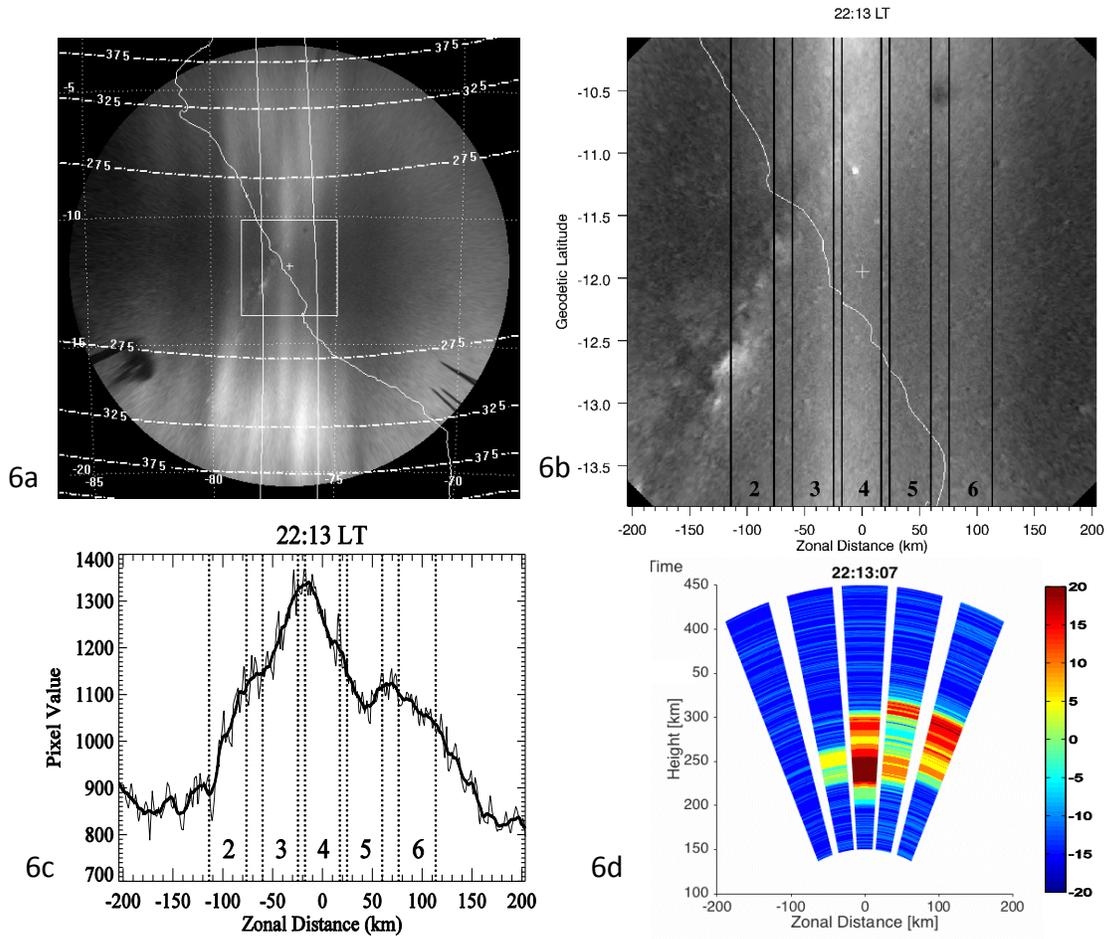

Figure 6

6a shows an unwarped image from 22:13 LT. Dashed lines showing the apex altitude of the magnetic field. Magnetic longitudes correspond to the edge of the field of view of AMISR-14 are shown as vertical white lines. Geographic latitude and longitude are marked by dotted lines. 6b is the area within the box in 6a with zonal distance away from the ASI as the x-axis and geodetic latitude as the y-axis. The positions of the AMISR-14 beams are shown, although in reality they do not cover the full N-S extent shown. 6c is a zonal cut through the center showing pixel value as a function distance. The AMISR-14 beams are also marked here. The lighter line is a 7 pixel meridional average and the darker line is a 15 pixel running average. Beam 2 is on the eastern wall of depletion "A". The data from AMISR-14 is shown in 6d. SNR is show in dB.

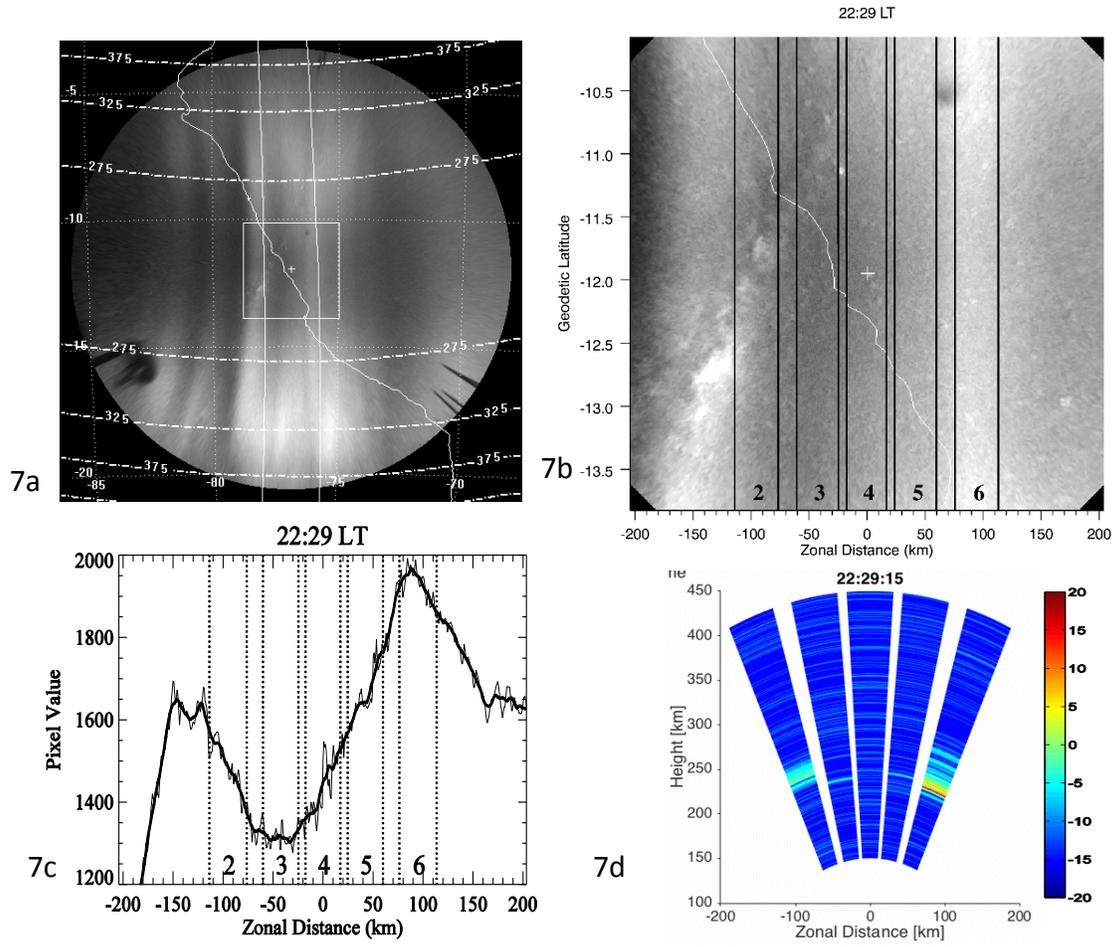

Figure 7

Same as Figure 5 but for a later time. Beam 3 is sampling the center of depletion "A".

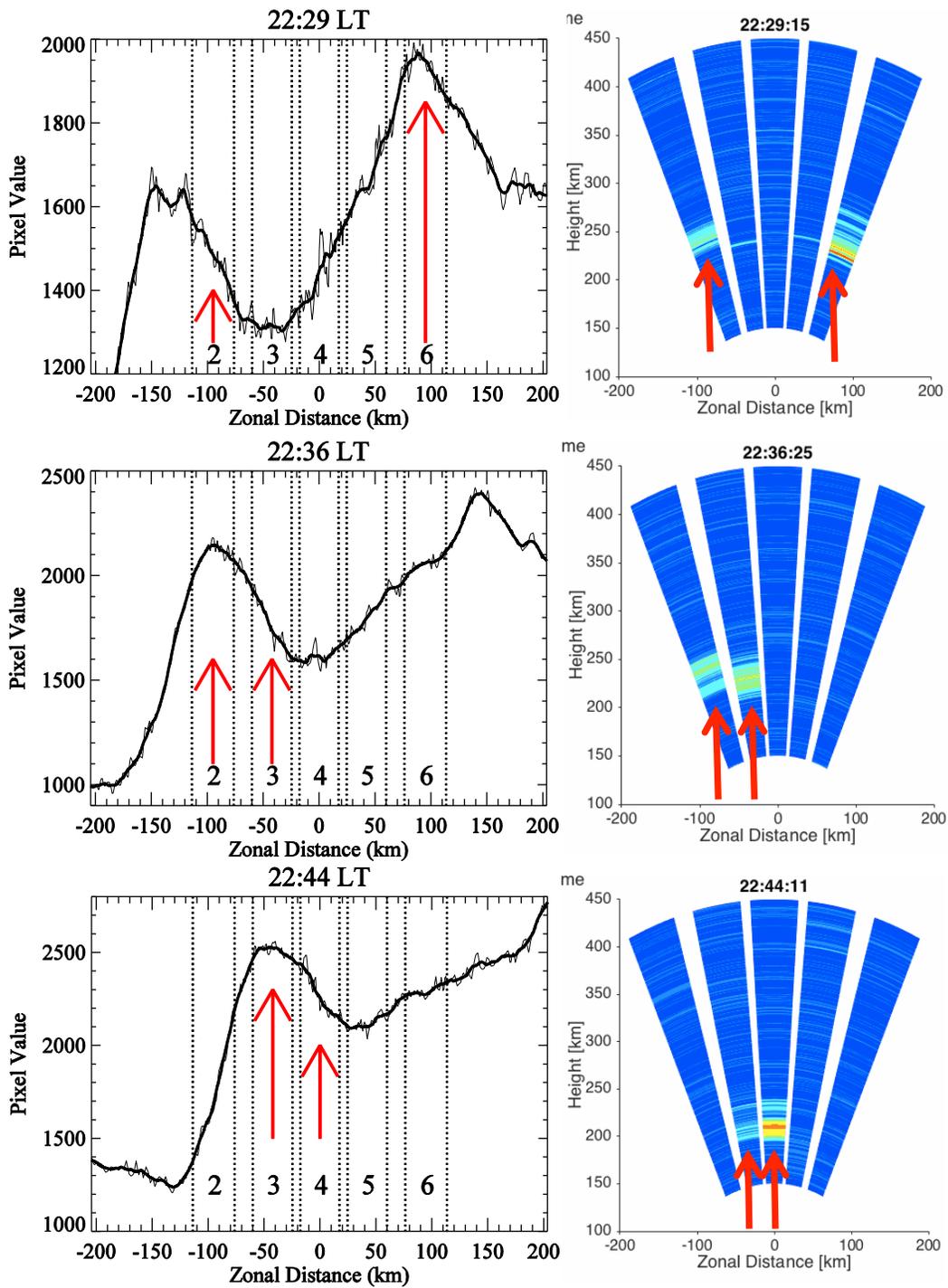

Figure 8

On the left are zonal cuts, as described in previous figures, for three consecutive images from 22:29 LT to 22:44 LT. Red arrows indicate where the echoes are occurring. The echoes in this example are only coming from the western wall of depletion "A".

35